\documentclass[reprint,prb,aps,showpacs,twocolumn]{revtex4}
\usepackage{epsfig}
\usepackage{amsmath}
\usepackage{dcolumn}

\begin{document}

\title{Jahn-Teller Distortion and  Ferromagnetism in the Dilute Magnetic
Semiconductors GaN:Mn }
\author{Xuan Luo and Richard M. Martin}
\address{Department of Physics, University of Illinois at Urbana-Champaign, 1110 W. Green street, Urbana, IL 61801}
\date{\today}
\draft    
\begin{abstract}
Using first-principles total-energy methods, we investigate
Jahn-Teller distortions in III-V dilute magnetic semiconductors,
GaAs:Mn and GaN:Mn
in the cubic zinc blende structure. The results for an isolated Mn
impurity on a Ga site show that there is no appreciable effect in
GaAs, whereas, in GaN there is a Jahn-Teller effect in which the
symmetry around the impurity changes from T$_{d}$ to D$_{2d}$ or
to C$_{2v}$. The large effect in GaN occurs because of the
localized d$^4$ character, which is further enhanced by the
distortion.
The lower symmetry should be detectable experimentally in cubic
GaN with low Mn concentration, and should be affected by charge
compensation (reductions of holes and conversion of Mn ions to
d$^5$ with no Jahn-Teller effect).
Jahn-Teller effect is greatly reduced because the symmetry at each
Mn site is lowered due to the Mn-Mn interaction. The tendency
toward ferromagnetism is found to be stronger in GaN:Mn than in
GaAs:Mn and to be only slightly reduced by charge compensation.

\end{abstract}
\vskip 12pt
\pacs{PACS numbers: 75.50.Pp, 76.50.+g, 72.25.Dc}

\maketitle

\begin{center}I. Introduction\end{center}
Electronics based upon the spin of the electron (spintronics)
seeks to exploit the spin of charge carriers in
semiconductors\cite{gnlref1,gnlref2,gnlref5}. It is widely
expected that new functionalities for electronics and photonics
can be derived if the injection, transfer and detection of carrier
spin can be controlled above room temperature in these dilute
magnetic semiconductors (DMS)\cite{gnlref3}. Most of the work in
the past has focused on InAs:Mn\cite{inasmn1,inasmn2,IN1,IN2},
GaAs:Mn\cite{gaasmndft1,gaasmndft2,gaasmndft3,gaasmndft4,gaasmndft5,
gaasmnmd1,gaasmnmd2,gaasmnmd3,gaasmnmd4,gaasmnmd5,gaasmnexp1,gaasmnexp2,
gaasmnexp3,gaasmnexp4,gaasmnexp5,gaasmnexp6,gaasmnexp7} and
Ge:Mn\cite{Ge1,Ge2}. Novel control of magnetism has already been
achieved in these host materials. However, the reported Curie
temperatures\cite{gnlref1} are too low to have significant
practical impact. Recently, there has been interest in wide
bandgap semiconductors, such as GaN\cite{gnlref4}, which may
exhibit higher
Curie temperatures\cite{Dietl,ganmndft1,ganmndft2,ganmndft3,ganmnmd,ganmnexp1,ganmnexp2}. Since there
has been tremendous progress on the growth of high-quality (Ga,
Mn)N epitaxial layers\cite{cubgan1,cubgan2}, GaN:Mn is a promising high Tc
ferromagnetic semiconductor. Using the Zener model of
ferromagnetism\cite{lochole}, Dietl et al.\cite{Dietl} predicted
that cubic GaN doped with  5 at.\% of Mn and containing a high
concentration of holes (3.5$\times 10^{20} cm^{-3}$) should
exhibit a Curie temperature exceeding room temperature. However,
the mechanisms of ferromagnetism in DMS materials is still an open
question.

Theoretically, there are many proposals for the electronic
configuration
which have focussed the ferromagnetic mechanism upon
the coupling between the host p and Mn 3d states\cite{d5hgaas,d4gan,d4d5rev,ploog,zhang-rice};
suggestions include Mn 3d$^{5}$+hole induced ferromagnetism in
GaAs:Mn\cite{d5hgaas}, whereas in  GaN:Mn the spin-spin
interaction has been proposed to be driven by a double exchange
mechanism involving of d-electrons in Mn 3d$^{4}$
states\cite{d4gan}.
It has been suggested that 3d$^{5}$, 3d$^{5}$+hole, and 3d$^{4}$
coexist in GaAs:Mn based upon evidence in the dilute regime that
there is a Jahn-Teller\cite{jt} distortion associated with the
3d$^{4}$ state\cite{ploog}.
However, to our knowledge only one work has proposed that there
should be a strong Jahn-Teller effect in the GaN:Mn
system\cite{zhang-rice}, and there have been no quantitative
theoretical studies of the Jahn-Teller effect in
either GaAs:Mn or GaN:Mn. In this paper, we report first
principles calculations of the magnitude of the Jahn-Teller effect
and the consequences for ferromagnetism in GaAs:Mn and GaN:Mn.

The Jahn-Teller effect was first proposed to occur in open-shell
molecules\cite{jt}, and there are many examples in impurity states
in II-VI\cite{jt-26} and III-V\cite{jt-35} semiconductors. The
effect is caused by a distortion that lowers the symmetry and
leads to a splitting of a degenerate state that is linear in the
magnitude of the distortion (Fig.1 (a)). If the state is partially
occupied, the total energy is always lowered by some distortion
since all other contributions to the energy are quadratic in the
distortion (Fig.1(b)),
and the total energy is minimum in a distorted configuration
(Fig.1 (c)). It may also happen that a large distortion occurs
leading to a new bonding configuration separated by an energy
barrier, such as the DX center\cite{jt-DX} and AX
center\cite{jt-AX}. Although the existence of a Jahn-Teller effect
is determined only by symmetry, we have to do a quantitative
calculation of the magnitude. If the splitting is less $\approx$
0.01 eV, we estimate that the effect is smaller than the effect of
quantum fluctuation or temperature and therefore can be ignored.

Our calculations were performed using the density-functional
theory within the generalized gradient approximation (GGA)
of the Perdew-Wang 91 form\cite{GGA}. We used the Vanderbilt
ultrasoft pseudopotentials\cite{vanderbilt} and the
Vosko-Wilk-Nusair interpolation for the correlation functional in
the spin-polarized calculations, as implemented by the plane-wave
total energy VASP code\cite{VASP}.
The calculated lattice constants are $a$ = 5.751 \AA\ for GaAs and
$a$ = 4.542 \AA\ for GaN, and all calculations with supercells are
done keeping the supercell lattice vectors fixed as multiples of
the primitive lattice vectors. All cell-internal structural
parameters are fully relaxed until the forces are converged to
within 0.05 eV/\AA. The cutoff energy for the planewave expansion
is 170 eV for calculations involving Mn in GaAs and 270 eV for Mn
in GaN, with check using 400 eV.
For self-consistent total energy
calculations, we used a 64-atom supercell and a 4x4x4
Monkhorst-Pack $k$-point mesh (which has been verified to be
sufficient\cite{gp3}), and selected cases were checked with a
6x6x6 $k$-point mesh. Density of states plots were made using a
finer 8x8x8 $k$-point mesh.

\section{Isolated Mn impurites}

In order to study isolated Mn impurites, we have carried out
calculations on 64 atom cells with one Mn substituted for a Ga
atom in GaAs and in GaN.
We have calculated the total energy and the eigenvalues of the
Kohn-Sham hamiltonian for various cases. First we have considered
the ideal geometry with all atoms at the ideal positions of the
GaAs or GaN tetrahedral lattices, and breathing distortions in which the bond
lengths change but the symmetry is constrained to remain T$_{d}$.
We have also considered three different lower symmetry distortions
as shown in Fig. 2: (a) T$_{d}$ to D$_{2d}$ (b) T$_{d}$ to
C$_{2v}$, (c) T$_{d}$ to C$_{3v}$. All cases can be compared if we
define symmetry-adapted variables. In order to disentangle the
effects of the distortions, we define the positions of the four
nearest-neighbor N atoms around a Mn atoms to be $\bf{R}_i^0$,
$i=1,4$ in the ideal tetrahedral structure, and the displacements
of the neighbors relative to the Mn, as:
\begin{equation}
\Delta{\bf{R}}_i=(\Delta{x}_i, \Delta{y}_i, \Delta{z}_i)
\end{equation}
The displacement of the neighbors can be projected into a two
parts.  One part is a symmetric radial ``breathing'' component,
which preserves the tetrahedral symmetry, given by
\begin{equation}
\Delta{\bf{R}}^B_i = \Delta{R}^B \; \hat{\bf{R}}_i^0,
\end{equation}
where the magnitude $\Delta{R}^B$ is easily extracted for a given
displacement pattern $\Delta{\bf{R}}_i$ using
\begin{equation}
\Delta{R}^B = \frac{1}{4} \sum_i \hat{\bf{R}}_i^0 \cdot
\Delta{\bf{R}}_i.
\end{equation}
The remaining parts of the displacement are symmetry-breaking
Jahn-Teller distortions, given by
\begin{equation}
\Delta{\bf{R}}^{JT}_i = \Delta{\bf{R}}_i - \Delta{\bf{R}}^B_i.
\end{equation}

In one set of calculations, we have considered only breathing,
varying the magnitude  of $\Delta{R}_i^B$ and forcing the symmetry
to remain Td.
Comparing to ideal tetrahedral positions,
for GaN all four N nearest neighbors of the Mn atom
move closer to the Mn by the amount $\Delta{R}_i^B = 0.034$ \AA\
and the energy decreases by $\Delta E = 0.05$ eV. For GaAs all
four As nearest neighbors of the Mn atom move closer to the Mn by
only a small amount $\Delta{R}_i^B = 0.007$ \AA\ and the energy
decreases by $\Delta E = 0.008$\ eV.
The difference between GaAs and GaN is readily explained by the
different size of As and N atoms.

In the remaining calculations the atoms are allowed to distort and
lower the symmetry. In the case of GaAs the energy was never
found to decrease. Therefore, we conclude that there is no
appreciable Jahn-Teller effect in GaAs.  However, in GaN the
energy is found to decrease substantially. The decrease in the
total energy for the three different symmetries is listed in Table
I. There we have defined the energy relative to the minimum energy
of T$_{d}$ symmetry as described above.
In the distortion shown in Fig. 2 (a), the four nearest neighbors
of Mn atom move to lower the symmetry from T$_{d}$ to D$_{2d}$
with displacements ($|\Delta{x_i}| = |\Delta{y_i}| \not=
|\Delta{z_i}|$).
The distortion results in a 0.10 eV lower energy
in GaN:Mn (see Table I the D$_{2d}$ column).
(This energy difference changes by only 6 meV if the energy cutoff
is increased from 270 eV to 400 eV). In the Fig.2 (b), two
neighbors of Mn atom move along [110] and another two neighbors of
Mn atom move along [\={1}10] direction. This atomic configuration
with 0.08 eV total energy lowering in GaN:Mn (Table I, C$_{2v}$
column). However, there is not an obvious effect for Fig. 2 (c)
distortion,
in which Mn atom move along [111] direction, as shown in the
column for C$_{3v}$ in Table I.

From these calculations, we conclude that a strong Jahn-Teller
effect should be observed for low concentrations in GaN (for D$_{2d}$ and C$_{2v}$)
but not in GaAs:Mn. Furthermore in GaN:Mn the distortion is
slightly favored for D$_{2d}$ symmetry; however, the energies are
close enough that either distortion may occur in actual systems.
We note that the lowering of energy from breathing relaxation 
alone ($0.05$ eV) is smaller than the lowering of the energy ($0.10$ eV and
$0.08$ eV relative to breathing relaxation) for the Jahn-Teller
distortion (both D$_{2d}$ and C$_{2v}$). This shows clearly the importance of the
Jahn-Teller effect. In addition we have analyzed the magnitudes
of the components of the D$_{2d}$ displacement for the lowest energy.
The magnitude of the breathing component $\Delta{R}^B$ is $0.020$ \AA\
and the magnitude of the Jahn-Teller displacement is
$|\Delta{\bf{R}}^{JT}_i| = 0.068$ \AA\
for each of the four
neighbors. Thus the magnitude of the components of the
displacement, also show that the Jahn-Teller effect,
is larger than the breathing for Mn in GaN.

We interpret as a Jahn-Teller distortion caused by the partial
occupation of the t$_{2d}$ state which is split due to the lower
$D_{2d}$ symmetry. The Mn triplet t$_{2d}$ of 3d energy level splits into
singlet and doublet states at $\Gamma$ point. The splitting
increases approximately linearly with the magnitude of the atomic
displacements, as expected. 
At the minimum energy position, the splitting energy $\Delta{E}_{t_{2d}}$ is 0.23 eV.
We have studied the stability of the
distorted state by varying the magnitude of the distortion in various ways.

We also considered large distortions in Fig.2 (b) and (c) that might lead to
qualitative rebonding of the atoms, e.g, two N (or As) atoms
forming nearest neighbors as in an AX center (like Fig.2 (b)) or large
displacements of the Mn atom along the [111] direction as in a DX
center (like Fig.2 (c)). However, no configuration was found to be stable or even
metastable state in either GaAs:Mn and GaN:Mn.

Figure 3 shows the partial density of states (DOS) of the Mn 3d
states in GaAs:Mn (Fig. 3 (a)) and cubic GaN:Mn in T$_{d}$
symmetry (Fig. 3 (b)) and the Jahn-Teller distortion of D$_{2d}$
symmetry (Fig. 3 (c)). The results for the undistorted cases (Fig.
3 (a) and 3 (b)) are very similar to those found by other
groups\cite{1,gp2,gp4,gp5}. In GaAs:Mn, we find
that Mn 3d states is 2 eV below host valence band minimum (VBM)
and the t$_{2d}$ level is lower than the e$_{d}$ level (see Fig. 3
(a)). On the other hand, in GaN:Mn, the Mn 3d state is above host
VBM and the t$_{2d}$ level is higher than e$_{d}$ (see Fig. 3
(b)). Because of this difference between GaAs:Mn and GaN:Mn, there
are different consequences for the Mn 3d states when the symmetry
changed from T$_{d}$ to D$_{2d}$: the t$_{2d}$ energy level has no
splitting ($<$ 0.01 eV) in GaAs:Mn, but a large splitting (0.23 eV
maximum at $\Gamma$ point) in GaN:Mn, with the Fermi level in the
gap as shown in Fig.3 (c).

For the case of an isolated Mn substituted for a Ga atom in GaN,
we also considered the charged state with an added electron.  This
charge state is expected when the system is compensated, reducing
the number of holes and leading to Mn atoms in the 3d$^5$ state.
Since this is a symmetric closed-shell state, the Jahn-Teller
distortion should disappear. Indeed, our calculations bear this
out. We carried out calculations in which the atoms are
constrained to have T$_{d}$ symmetry (pure breathing relaxation)
and other calculations in which the atom positions are relaxed
starting with distorted initial positions having D$_{2d}$
symmetry. In all cases the total energy is the same to within 0.02
eV, the splitting of t$_{2d}$ is very small (about 0.01 eV), and
the atoms relax to positions near T$_{d}$ symmetry . In principle,
all calculation should have the same total energy. We consider the
differences to be negligible and these results can be considered
as numerical tests showing the accuracy of our calculations.
Clearly, this supports our conclusion that an isolated neutral Mn
substituted for Ga has a Jahn-Teller distorted ground state in
GaN, and that compensation can decrease the effect causing some Mn
to in the 3d$^5$ state with no Jahn-Teller distortion.

\section{Interactions of Mn-Mn pairs}

Although we have shown that isolated Mn impurities undergo a large
Jahn-Teller distortion in GaN, further studies are required to
establish the effects, if any, upon the properties of GaN:Mn
alloys in interesting concentration ranges for magnetic
semiconductors. Since ferromagnetism is due to interactions
between the Mn atoms, we must consider the effect upon Mn-Mn
interactions for Mn atoms separated by typical distances found in
the alloys. The Jahn-Teller effect would be expected to cause a
reduction in the tendency toward ferromagnetism since it leads to
a splitting of the states. The splitting will reduce the
interactions between Mn pairs since hoping requires an extra
energy cost. On the other hand, the interaction between the Mn
atoms may be so large that it dominates over the Jahn-Teller
effect.

In order to study Mn-Mn interactions, we have carried out
calculations on 64 atom cells with two substitutional Mn atoms at
various distances and in different spin states. We selected 2 Mn
atoms separated by distances $\sqrt{2} a$ and $\sqrt{2} a/2$,
where $a$ is the cubic GaAs or GaN lattice constant. For GaAs
case, for separation $\sqrt{2} a$, the total energy difference
between ferro- and antiferro-magnetic spin states $\Delta{E}_{AF}$=0.22
eV/Mn-pair, which is comparable with 0.2 eV/Mn-pair in
Ref.\cite{1} and 114 meV/Mn (0.228 eV/Mn-pair) in
Ref.\cite{gp3}. For the GaN case, we have considered different
starting configurations of the atoms, in one set of calculations
starting with all atoms in the ideal zinc blende positions and in
a second set starting with Jahn-Teller distorted states (the atoms
around the Mn-Mn pairs are placed in distorted configurations).
The final positions of the atoms after relaxation are the same for
the two cases, showing that checked that final configuration
was independent of the starting point. For separation $\sqrt{2}
a/2$, the energy difference is $\Delta{E}_{AF}$=0.36 eV/Mn-pair,
this value is good in comparing to similar work\cite{2,gp3},
which is 188 meV/Mn and 156 meV/Mn. For separation $\sqrt{2} a$,
the $\Delta{E}_{AF}$=0.30 eV/Mn-pair\cite{Footnote}, while a similar work
\cite{gp2} gave 161 meV/Mn.

The energy levels for the Mn 3d states in the gap show that the
interactions between the two Mn atoms in a pair is indeed larger
than the splitting caused by the Jahn-Teller effect on the
individual Mn atoms. In the ferromagnetic state the d energy
levels are split by the Mn-Mn interactions by amounts that are
much larger than the splitting due to the Jahn-Teller effect.  The
maximum width of the d-bands in our supercell calculation is 0.4
eV for the Jahn-Teller splitting as shown in Fig. 3 (c). In contrast,
the maximum width of the d-bands is 1.4 eV for a ferromagnetic
Mn-Mn pair separated by $\sqrt{2}$a/2 with all atoms relaxed.
Furthermore, the Mn-Mn interactions are present even if all atoms
are in ideal positions, in which case the maximum width is only
slightly changed to 1.3 eV. Similar results are found for Mn-Mn
pairs separated by $\sqrt{2}$a. These results show that
Jahn-Teller distortions do not have a large affect upon the
magnetic interactions between Mn atoms at distances expected in
actual ferromagnetic semiconductors.

It is interesting also to consider the realistic case with charge
compensation, in which some of the t$_{2d}$ levels are filled. If
the states are localized near the Mn atom as in GaN:Mn, each added
electron can be interpreted as a conversion of a d$^4$ into a
d$^5$ state. Since a d$^5$ state is a spatial singlet with no
degeneracy except spin, no Jahn-Teller effect will occur.
 We also
calculated -1 charged states for Mn pair separated with $\sqrt{2}$
a and $\sqrt{2} a/2$ GaN lattice constant. All the results as well
as neutral Mn-Mn pair case are listed in Table II. We found that
the total energy difference between ferro and anti-ferro magnetic states are
decreased in charged states, such as for $\sqrt{2} a/2$,
$\Delta{E}_{AF}$ changes from 0.36 eV/Mn-pair to 0.30 eV/Mn-pair ,
and for $\sqrt{2} a$ case, $\Delta{E}_{AF}$ changes from 0.30
eV/Mn-pair to 0.22 eV/Mn-pair (see Table II). This result show that charge
compensation will decrease the tendency for ferromagnetism, as
expected.

\section{Conclusions}

In summary, we have studied the Jahn-Teller distortion in GaAs:Mn
and GaN:Mn. Our results show that a strong Jahn-Teller distortion
should happen in uncompensated GaN:Mn
at low concentrations where the Mn impurities are isolated. The
lowering of the energy is due to the splitting of the t$_{2d}$
states of the localized d-electrons on the Mn 3d$^{4}$ ion,
leading to an energy gap. There are two possible symmetries
C$_{2v}$ and D$_{2d}$, with the latter having the lowest energy.
In the presence of charge compensation, the Mn d states are filled
leading to filled shell spherically-symmetric 3d$^{5}$ ions and
the Jahn-Teller effect disappears. These effects should be
observable experimentally.

In contrast, in GaAs:Mn the Mn 3d states are primarily 3d$^{5}$
with a hole in the GaAs valence band.  This state is only
weakly coupled to the distortions and the tendency for a
Jahn-Teller distortion is a negligible effect.

In order to study the effects upon magnetism, we carried out
calculations on Mn pairs. In agreement with other work, we find
Mn-Mn interactions to lead to ferromagnetism in both GaAs and GaN,
with larger interaction in GaN. The interaction between Mn-Mn
pairs at realistic distances is sufficiently large that it
dominates over the Jahn-Teller effect. The interactions between
Mn atoms is not greatly affected by lattice relaxations and there
is always a clear tendency for ferromagnetic alignment of Mn
pairs. In realistic cases with charge compensation, ferromagnetism
is still favored: even with 100\% compensation the ferromagnetic
interactions are reduced by only ~20\%.  Finally, even though our
calculations are for the cubic structure, the conclusions on
magnetic interactions should carry over to the wurtzite structure
since they do not depend upon detailed positions of the atoms and
the ferromagnetic state persists whether or not there are
distortions. Thus our results support the conclusion that GaN:Mn
holds promise as a ferromagnetic semiconductor.

This work was supported by the DARPA under contract
No.N0014-01-1-1062 and by the U.S. Department of Energy through
the Computational Materials Science Network (CMSN) and through
Grant No. DEFG02-91ER45439 to the Frederick Seitz Materials
Research Laboratory at the University of Illinois at
Urbana-Champaign. Computer time was provided by NCSA and the
Materials Computation Center at the University of Illinois at
Urbana-Champaign.  We are appreciative for helpful conversation
with S. B. Zhang, S. Chiesa, S. H .Wei.

\begin{figure}
\caption{ A schematic drawing of the Jahn-Teller distortion. (a)
Degenerate energy level splitting, which leads to a linear decrease in the energy. 
(b) Strain contribution, which increases the total energy quadratically.
(c) Formation of a stable configuration with an total energy minimum.
} \label{deformation structure}
\end{figure}

\begin{figure}
\caption{
A schematic drawing of three different atomic configurations, which lower
symmetries in Jahn-Teller distortion (a)T$_{d)}$ to D$_{2d}$, (b) T$_{d)}$ to C$_{2v}$,
(c)T$_{d)}$ to C$_{3v}$.
}
\label{deformation structure}
\end{figure}

\begin{figure}
\caption{ Partial density of states and schematic energy level
diagrams of Mn 3d states in (a) GaAs:Mn, (b) GaN:Mn in T$_{d}$ symmetry without a Jahn-Teller
distortion (no energy level splitting), (c) With Jahn-Teller distortion of D$_{2d}$ symmetry,
which shows the splitting of the t$_{2d}$ d-states.}   
\label{energy level splitting}
\end{figure}

\begin{table}
\caption{Comparison of total energy among four different symmetries in GaN:Mn. Here
the total energy ${E}_{tot}$ of
$T_{d}$ symmetry (with breathing relaxation) is set to zero.}
\begin{center}\begin{tabular}{ccccc}
\hline
\hline
Symmetry  & T$_{d}$    & D$_{2d}$    & C$_{2v}$   &  C$_{3v}$   \\
\hline
${E}_{tot}$ (eV)   & 0.00   &  -0.10   & -0.08     & -0.02     \\
\hline
\hline
\end{tabular}\end{center}
\end{table}

\begin{table}
\caption{Total energy difference $\Delta{E}^{AF}$ between ferro- (FM) and antiferro- (AFM) magnetic spin states for a
Mn-Mn pair in neutral and -1 charged
GaN:Mn. Here $\Delta{E}^{AF}$=${E}_{tot}^{AFM}$-${E}_{tot}^{FM}$, $\sqrt{2} a/2$ and $\sqrt{2} a$ are two separations of 
the Mn-Mn pair, and $a$ is lattice constant of cubic GaN.}
\begin{center}\begin{tabular}{ccc}
\hline
\hline
System & {Neutral GaN:Mn}  & {(-1) charged GaN:Mn}   \\
\hline
$\Delta{E}_{\sqrt{2} a/2}^{AF}$ (eV/Mn-pair)  &  0.36 & 0.30      \\
$\Delta{E}_{\sqrt{2} a}^{AF}$ (eV/Mn-pair)    &  0.30 &  0.22     \\
\hline
\hline
\end{tabular}\end{center}
\end{table}


\begin{references}
\bibitem{Footnote} In our calculations, we found that care must be taken in the calculation, to ensure that one with correct spin-zero antiferro-magnetic states instead of a high-energy non-magnetic states.
\end{references}
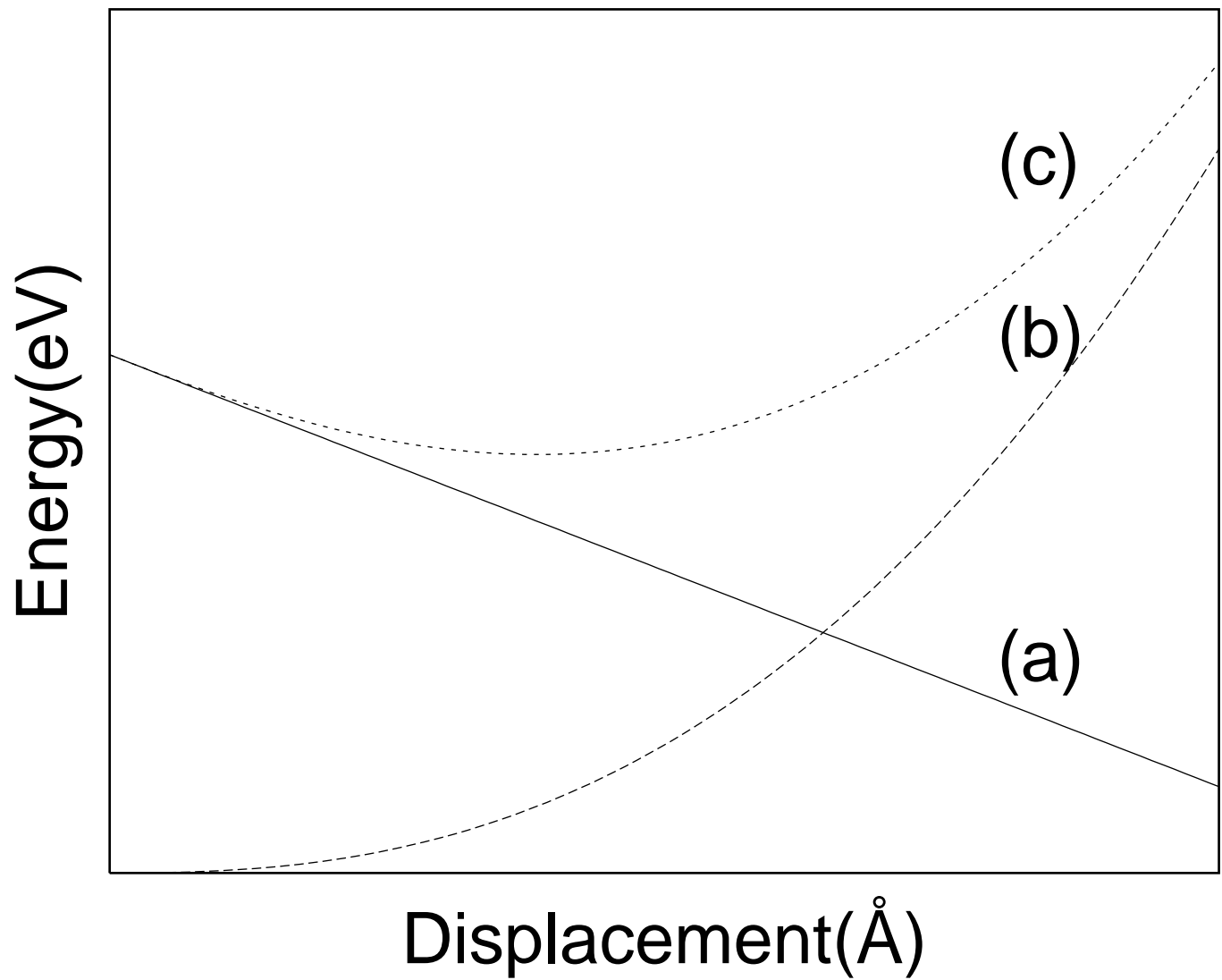
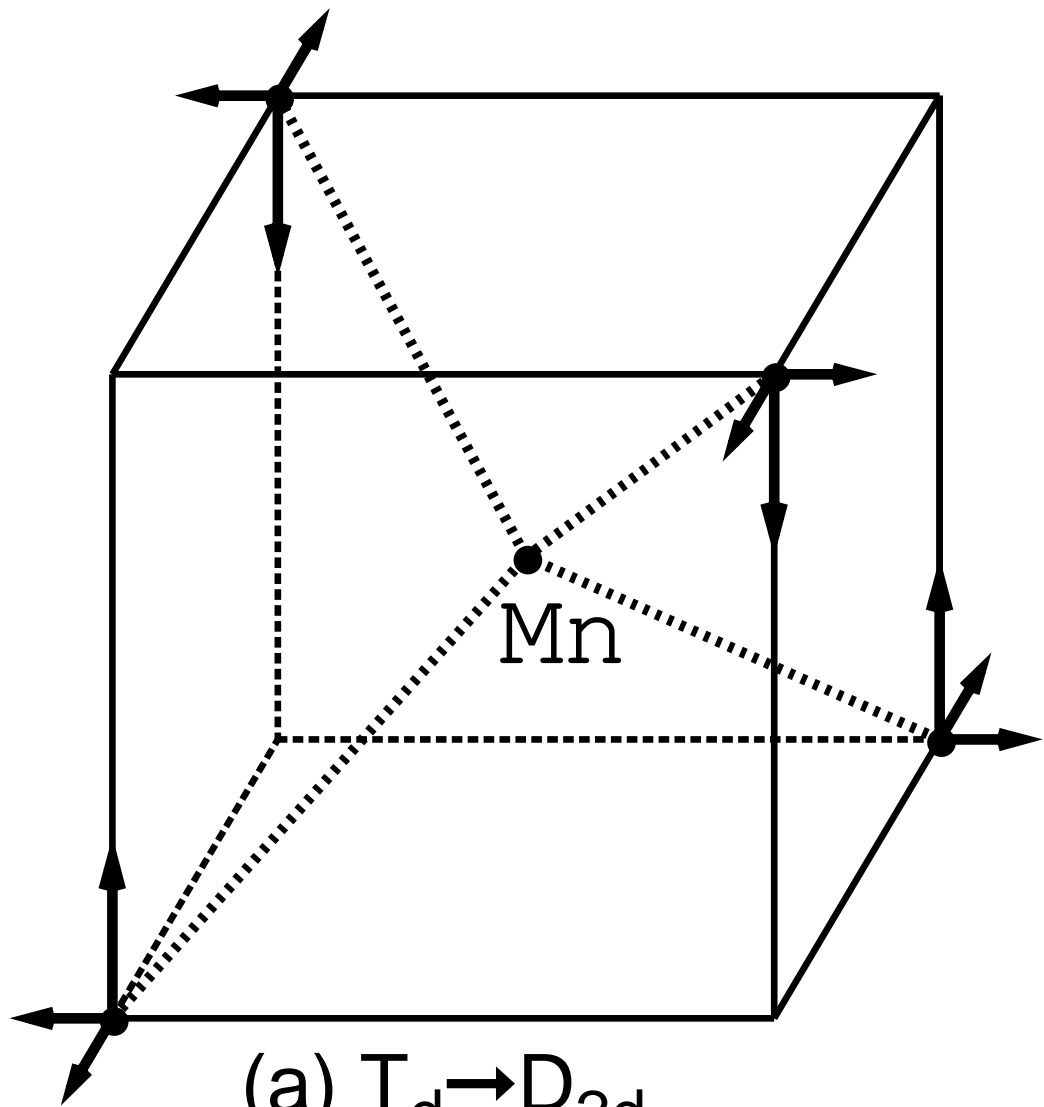
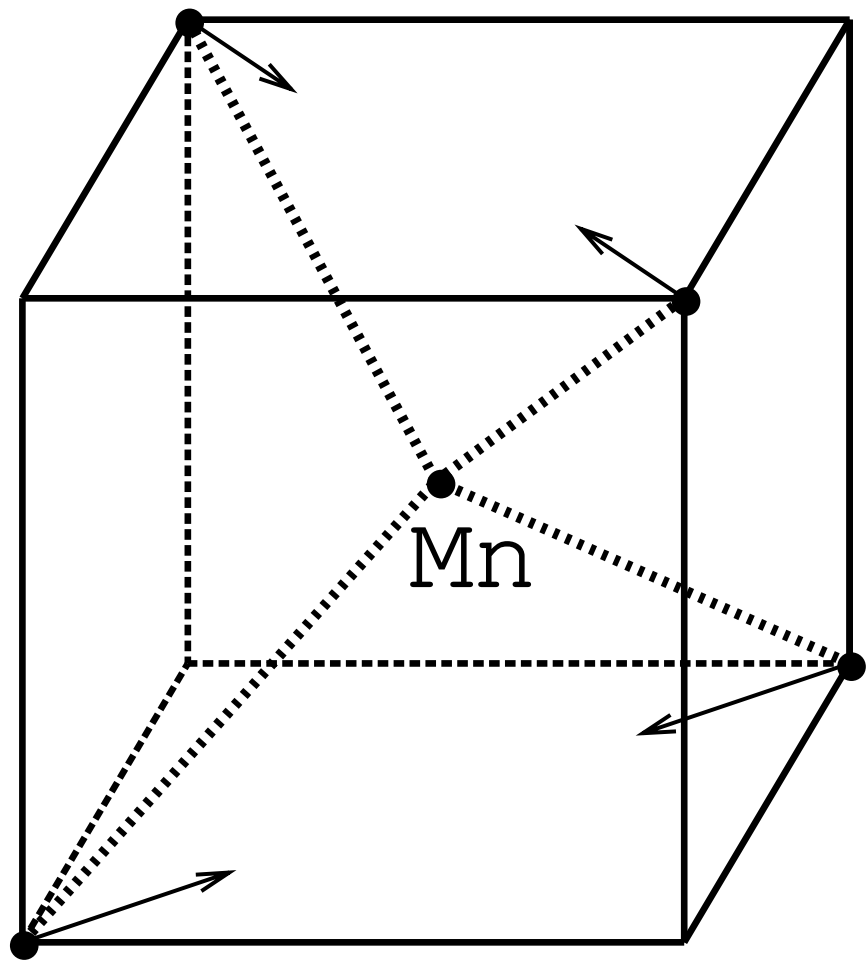
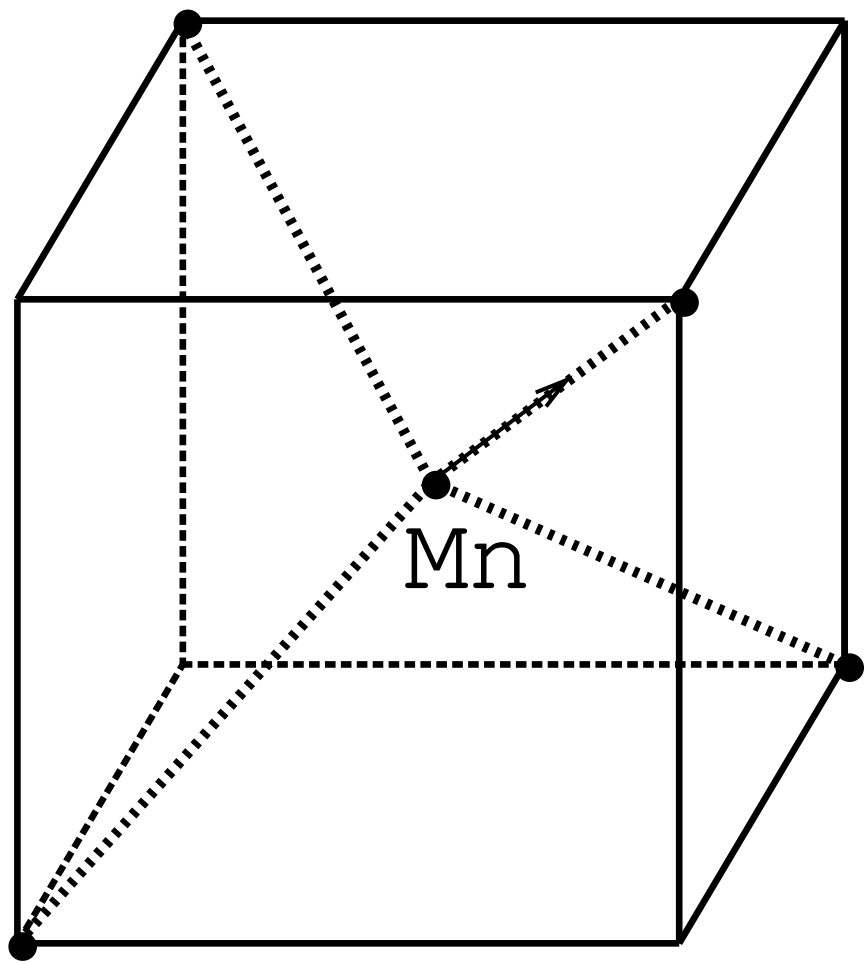
\end{document}